\documentstyle[12pt]{article}
\begin{document}

\title{ {\Large {\bf Superluminal Motion in The interacting 
System of Electrons, Positrons and Photons } } }

\author{\vspace{1cm}\\
{\small T. M. Gassym \thanks 
{e-mail: gassymt@newton.phycics.metu.edu.tr}} \\
{\small Physics Department, Middle East Technical University} \\
{\small 06531 Ankara, Turkey} \\
{\small Institute of Physics Academy of Sciences of Azerbaijan
Republic, Baku} } 

\date{}

\begin{titlepage}
\maketitle
\thispagestyle{empty}

\begin{abstract}
The coupled system of Boltzman equations for the interacting system of 
electrons, positrons and photons in high external electric ,$\vec{E}$, 
and arbitrary magnetic ,$\vec{H}$, fields is solved. The consideration 
is made under the conditions of arbitrary heating and the mutual drag 
of carriers and photons.

The non--stationary and non--uniform distribution function of
photons 
for the all considered cases is obtained. It is shown that the 
distribution function of photons has the stationary limit for the
drift velocities $\left(\vec{u}\vec{q}/ \hbar \omega_q\right) <1 $ 
and grows exponentially with time for the drift velocities 
$\left(\vec{u}\vec{q}/ \hbar \omega_q\right) \geq 1$. It is also 
shown that the mutual drag of carriers and photons leads to the 
formation of ``quasi-particles''--the ``electron dressed by photons'' 
and ``positron dressed by photons'', i.e. the mutual drag plays the 
role of the dressing mechanism of carriers and leads to the 
renormalization of the mass and frequency of photons.
 
As a result of the analyses of the phenomena of coupling of the mutual 
drag system of carriers with the photons, we obtained some fundamental
results: {\bf a)} the finiteness of the mass of photon (i.e. the rest 
mass of the photons is not zero); {\bf b)} reality of tachyons as a 
quasi--particles with a real mass in amplifying system (or regime); 
{\bf c)} the mechanism of the theory of relativity and that of the 
Doppler effect which coincides with the renormalization of frequency 
or mass as a result of the mutual drag of carriers and photons
at external field (force) are the same. These conclusions were obtained 
as a result of the fact that the relativistic factor enters the
expressions of the distribution function of photons and other physical
expressions in the first order in the form $\left[1-\left(u^2/c^2\right) 
\right]^{-1}$, but not in the form $\left[1-\left( u^2/c^2\right) 
\right]^{-1/2}$ as in Einstein's theory. Moreover, it is shown that the
time delatation which is relativistic effect really takes place for the
relaxation time of carriers or for the life time of carriers, or for the
period of electromagnetic oscillations. The latter is a direct result of
the Doppler effect. Also it is shown that the velocity of light is an
average velocity of photons in the ground state, like the velocity of
sound is the average velocity for the phonons.
\end{abstract}
\end{titlepage}

\section{Introduction}

In my earlier publications I made a theoretical investigation of 
interacting system of electrons and phonons in semiconductors, 
semimetals and gaseous plasmas under high external electric and 
magnetic fields, and also when a strong electromagnetic wave 
propagating in the materials. In [1-4] the coupled system of kinetic
equations for interacting electrons, holes and phonons under high 
electric and magnetic fields were solved by taking into account the
arbitrary heating and the mutual drag of carriers and phonons. For the
phonons, the solution of non--stationary kinetic equation was found and 
it was shown that the non--equilibrium and non-stationary distribution
function of phonons has the stationary limit, when the drift velocity 
of coupled by the mutual drag  system of carriers and phonons is less 
than the velocity of sound $(u<s)$. For the drift velocities $u>s$, 
the distribution function of phonons grows exponentially with time. 
It corresponds to the generation of intrinsic phonons and amplification 
of phonons introduced to the system externally (the stream of phonons).

It was shown that the mutual drag leads to the renormalization of the 
frequency or the mass of carriers. As a result of the mutual drag,
electrons and holes are ``dressed'' by phonons and forms 
``quasi--particles'' which have the electron or hole charge 
$(\pm e)$ and the phonon mass $m \simeq T_i/s^2$, where $T_i$ is the 
temperature of the coupled mutual drag system of carriers and 
phonons)\cite{1}--\cite{5}. In weak electric fields, the mass 
of phonons is $m_0 \simeq T/s^2$. In the case of propagation of 
strong electromagnetic waves in semiconductors and semimetals 
under the external magnetic field, this leads to the cyclotron 
resonance of the ``quasi--particles'' with frequencies 
$\omega_H=(3/4)e H/ \left[(T_i/s^2)c\right]$\cite{5,6}. It was
shown\cite{7} that the coupling of carriers and phonons as a result 
of their mutual drag is a general phenomenon. Actually, as it follows
from [7], under the condition of longitudinal propagation of a strong
electromagnetic wave in semimetals having equal electron and hole   
concentrations, the frequencies of Alfven and magneto--sound waves 
are equal to each other and have the form:

\begin{equation}
\omega_A=\omega_{ms}=kH/(4mN)^{1/2}=k v_A, 
\end{equation}
where $m=4(T_i/s^2)/3$ is also so--called ``mass of phonons''. 
$T_i=T_i(+0)$--the effective temperature of the coupled by the mutual 
drag system of ``electron + phonons'' or ``hole + phonons''. Thus we 
may conclude that the mutual drag of carriers and phonons under external
fields leads to the formation of compound particles (``quasi--particles'', 
the carriers dressed by phonons) a with drift velocity $\vec{u}$.

In the first investigation\cite{8}, the degenerate semiconductors and 
semimetals in which weak electromagnetic waves propagates were
considered. Particularly, in [8] the degenerate semiconductors having 
only one type of carriers were studied and the cyclotron resonance
of the coupled system of electron + phonon was predicted. But
soon after this work, it was shown\cite{9} that this resonance is not
observable because of the presence of impurities with a concentration
$N_i=n$ (where $n$ is the concentration of electrons). Also in the same 
work it was shown that in pure semimetals it is necessary to take into
account the presence of two types of carriers which are drifted in 
opposite directions at $H=0$. The remarks made on non--observability 
of the resonance in [9] applies also to the uniform and non--uniform 
low frequency cyclotron resonance in the degenerate semiconductors and 
in the semimetals with only one type of carriers when a strong
electromagnetic wave propagates in the materials\cite{5,6,10}.
These problems were discussed in\cite{3,5} and it was shown that in
intrinsic semiconductors and semimetals both uniform and non--uniform
cyclotron resonances in quasi--particle ``hole + phonon'' take place. 
In [3,5] it was also shown that such type of resonances are realized in 
non--degenerate impurity and intrinsic semiconductors and the question
on their observability was discussed.

\section{General Considerations}

In the present paper the coupled system of kinetic equations for 
interacting system of electrons, positrons and photons in external 
high electric field $\vec{E}$ and arbitrary magnetic field $\vec{H}$ 
is solved. The non--equilibrium distribution functions of electrons, 
positrons and photons are obtained by taking into account their 
arbitrary heating  and mutual drag.

For photons, the general solutions of non--stationary and 
non--uniform Boltzmann equations were found. The cases of weak 
$(\omega_H^{\pm}\tau_c \ll 1)$, classically high 
$(\omega_H^{\pm}\tau_c \gg 1)$, and quantizing 
$(\hbar\omega_H^{\pm} \gg T,T_c)$ magnetic fields were considered. 
Here $T$ is the initial temperature of equilibrium system (at $t=0$, 
before the external fields are applied), $T_c(\vec{E},\vec{H})$ is 
the temperature of heated electrons and positrons, $\omega_H^{\pm}$ is
their cyclotron frequencies, and $\tau_c^{-1}=\nu_c$ is the relaxation 
frequencies of electrons and positrons (the photon produced by the
annihilation of electron--positron pairs or by the scattering of 
electrons and positrons by photons). Also note that the index $c$ 
stands for electrons,(e), and positrons,(p).

In the absence of or in a weak magnetic fields, the inter--carrier
collision frequencies $\nu_{ee}$ and $\nu_{pp}$ are assumed to be 
much greater than the others, and that is why the isotropic parts 
of the distribution functions of carriers are assumed to be 
equilibrium one, with effective temperature of carriers
$T_c=T_{e,p}(\vec{E},\vec{H})$. This approximation corresponds to 
the case of high carrier concentrations, $n>n_{cr}$.

If the external field is a strong electromagnetic wave, then 
it is necessary to satisfy the following condition also:

\begin{equation}
\omega \gg \nu_\varepsilon^{\pm}
\end{equation}
where $\nu_{\varepsilon}^{\pm}$ is the collision frequencies of 
carriers for the energy transfer to scatterers and $\omega$ is the
frequency of the electromagnetic wave.

Under the condition (2), the isotropic parts of carrier distribution
functions with $(\nu_{\varepsilon}^{\pm}/ \omega)$ accuracy do not 
depend on time directly. By the direct solution of quantum kinetic
equations in general case for the arbitrary spherical symmetric 
dispersion law of carriers, it was shown that under quantizing and
classically high magnetic fields, the stationary distribution 
functions of carriers satisfying the boundary conditions, 
$F^{\pm}(\varepsilon)\mid_{\varepsilon \rightarrow \infty}=0$ 
have the form (for the arbitrary quantities of their concentrations):

\begin{equation}
F^{\pm}(\varepsilon)=\left\{const^{-1} exp \left(\int^{\varepsilon}
\frac{d\varepsilon'}{T_c(\varepsilon',t)} \right)+1\right\}^{-1},
\end{equation}
here $T_c(\varepsilon,t)=A(\varepsilon,t)/B(\varepsilon)$ is the
temperature of carriers which have occupied the energetic level 
$\varepsilon$ and

\begin{eqnarray}
A(\varepsilon,t)=(2\pi / \hbar)\sum_{\alpha \beta \vec{q}}
C_q^2 \mid I_{\alpha\beta}^2 \mid (\hbar \omega_q^*)^2
N(\vec{q},t)\delta(\varepsilon_{\beta} - \varepsilon_{\alpha} 
-\hbar\omega_q)\delta(\varepsilon_{\alpha} -\varepsilon), \\
\nonumber B(\varepsilon,t)=(2\pi / \hbar)\sum_{\alpha \beta \vec{q}}
C_q^2 \mid I_{\alpha\beta}^2\mid (\hbar \omega_q^*)^2
\delta(\varepsilon_{\beta} - \varepsilon_{\alpha} - 
\hbar\omega_q) \delta(\varepsilon_{\alpha} -\varepsilon), 
\end{eqnarray}
where $\hbar \omega_q^*=\hbar \omega_q -\vec{V}\vec{q}$,
$C_q$ is the constant of interaction and $I_{\alpha\beta}$ is a
matrix element for the transition from state $\alpha$ to state 
$\beta$ and back (reverse).

For arbitrary degree of quantization we have:

\begin{eqnarray}
F^{\pm}(\varepsilon)=\left\{1+\exp \left(\varepsilon - 
\zeta(\vec{E}\vec{H})/T_c \right) \right\}^{-1}, \\
\nonumber T_c=T_i\left\{ 1+\left[ \frac{V^{\pm}}{u}-1\right]^2 
(\varphi_1-1) \right\}, \\
\nonumber \varphi_1=\left[1-(\frac{u}{c})^2 \right]^{-1/2}. 
\end{eqnarray}
In the classical region of strong magnetic field we have:

\begin{eqnarray}
T_c=T_i\left\{1+\frac{1}{3}\left(\frac{V^{\pm}}{c}\right)^2
+\left(1-\frac{V^{\pm}}{u}\right)(\varphi_2-1) \right\}, \\
\nonumber \varphi_2=\frac{c}{2u}\ln \left| 
\frac{c+u}{c-u}\right|, 
\end{eqnarray}
here $V^{\pm}=cE/H$ is the Hall's drift velocity of carriers.

It was shown that for all cases considered the solution of 
non--stationary kinetic equation for the photons is:

\begin{eqnarray}
N(\vec{q},t)=\left\{N(\vec{q},\vec{r}-\overrightarrow{u_0} t,0)
+ \beta \int_0^t d\tau~ N(\vec{q},\tau') exp \left(-\int_0^\tau 
\gamma_q(\tau') d\tau' \right) \right\} \\
\nonumber exp\left(-\int_0^t \gamma_q(\tau) d\tau \right),
\end{eqnarray}
where $N(\vec{q},\vec{r}-\overrightarrow{u_0}t,0)$ is the initial
distribution function of photons in the absence of electric and 
magnetic fields (at $t=0$), which in the case of uniform space is the
equilibrium Planck's distribution function at temperature $T$. The 
increasing increment of photons is

\begin{equation}
\gamma_q=\beta \left(\frac{\vec{u}\vec{q}}{\hbar\omega_q}-1\right),  
\end{equation}
where $\beta=\beta_e+\beta_p+\beta_{ph}+\beta_b$ is the total 
collision frequency of photons with electrons, $(e)$, positrons, 
$(p)$, (including the photon decay to electron--positron pair), 
photons, $(ph)$, and boundaries, $(b)$, of the region occupied by 
the system, if such one exists.

\begin{equation}
\vec{u}(t)=\sum_{\pm}\overrightarrow{u^{\pm}}(t); 
~~\overrightarrow{u^+}(t)= \frac{\beta_e}{\beta}
\overrightarrow{V^+}(t); ~~\overrightarrow{u^-}(t)
=\frac{\beta_p}{\beta}\overrightarrow{V^-}(t), 
\end{equation}
where $\overrightarrow{V^{\pm}}(t)$ is the average drift velocity of 
the carriers, $\overrightarrow{u^-}(t)$ is the drift velocity of the
mutual drag system of ``electron + photons'' and $\vec{u}^{+}(t)$ is 
the same for the system of ``positron + photons''.

In general case, when the carriers are heated by a field of strong
electromagnetic wave $\vec{E}=\overrightarrow{E_0}
e^{-i\omega t}+\overrightarrow{E_0^{\ast}} e^{i \omega t}$,
$\overrightarrow{V^{\pm}}(t)=\overrightarrow{V^{\pm}} 
\cos\omega t$ we have 

\begin{eqnarray} 
N(\vec{q},t)=\left\{
N(\vec{q},0)+\beta \int_0^t d\tau N(\vec{q},\tau) 
exp \left(\beta \left[\tau-\frac{\vec{u}\vec{q}}
{\hbar \omega_q} \frac{\sin \omega \tau}{\omega}\right]
\right)\right\} \\
\nonumber exp \left( -\beta \left[t-\frac{\vec{u}\vec{q}}
{\hbar \omega_q}\frac{\sin \omega t}{\omega}\right]\right).
\end{eqnarray}
In the case of constant external electric field 
$(\omega \rightarrow 0)$ we have:

\begin{equation}  
N(\vec{q},t)= \left\{
N(\vec{q},0)-\frac{N(\vec{q},T_i)}{1-\vec{u}\vec{q}
/{\hbar \omega_q}}\right\} exp \left(\beta \left[
\frac{\vec{u}\vec{q}}{\hbar \omega_q}-1 \right]t \right)
+\frac{N(\vec{q},T_i)}{1-\vec{u}\vec{q}/{\hbar \omega_q}},
\end{equation}
here $T_i=(\beta_c/ \beta)T_c + (\beta_{ph}/ \beta)T_{ph}
+(\beta_b/ \beta)T_b$ is the temperature of the mutual drag 
system of coupled heated carriers and photons.
 
We have still considered the case when the initial state of the  
photons $(t=0)$ was assumed to be equilibrium state without any 
preferred direction. If the initial distribution part of the
photons has directional drift (the photon stream), then the 
solution of the kinetic equation for photons has the form:

\begin{eqnarray}
N(\vec{q},\vec{r},t)=\left\{
N(\vec{q},\vec{r}-\overrightarrow{u_0} t,0)+
\beta N_i(\vec{q},T_i) \int_0^t exp \left( 
\beta \int_0^\tau \left[1-\frac{\vec{u}\vec{q}}{\hbar \omega_q}
\right]d\tau' \right) d\tau \right\} \\
\nonumber exp \left(-\beta \int_0^t\left[1-
\frac{\vec{u}\vec{q}}{\hbar \omega_q} \right] d\tau \right).
\end{eqnarray}

As it follows from (12), the solutions of the uniform and 
non--uniform equations for the photons have the same form 
corresponding to the different initial conditions. Therefore, 
if the initial distribution function of photons has the form 
of the drifted Planck's distribution function
$N(\vec{q},\vec{r}-\overrightarrow{u_0}t,0)$ and if the external 
field is uniform, then this non--uniformity has to be preserved 
with time and the drift at external field has to be added to them. 
Thus the equation (12) allows us to consider the processes of 
absorption or amplification of photons introduced to the system 
from outside (initial stream of photons), and the generation of 
intrinsic photons of system in external field. In principle it is 
the most common form of the initial distribution function, which is  
taking the chance for examination of the affirmation of the special 
theory of relativity about equivalency of all the inertial frames 
of reference. Actually, by the transition to the frame of reference
drifting jointly with photons, as a result, we have the Planck's
equilibrium distribution function at the temperature $T$ in this 
frame of reference. In other words, in the absence of external 
fields $(E=H=0)$, from the kinetic equation, for initially 
non--uniform system of photons we obtain uniform one, by the 
transition from one frame of reference to another one but it does 
not mean that the two frames of reference are equivalent.

In fact, the transition from the frame of reference drifting jointly 
with the photons having velocity $\overrightarrow{u_0}=c\vec{q}/q$ to 
the frame of reference in which the photons are at rest is equivalent 
to the transition from one inertial frame of reference with $u=u_0)$ 
to the other with $u=0$. As a result we obtain 
$N(\vec{q},\vec{r}-\overrightarrow{u_0}t,0)=N(\vec{q},0)=N_0(\vec{q},T)$,
for all moments of time, solution (12) transforms to (11) but it does not 
mean that these two frames of reference are equivalent. In other words, 
the demand of equivalency of the laws of Physics in that two inertial
frames of reference is equivalent to the demand of the equivalency the
equilibrium Planck's distribution function to the drifted Planck's
distribution function or to the demand of equivalency of the
laws of Physics in the uniform and non--uniform cases (or spaces).

As one can see from (7), (11) and (12), the general solution of
non--stationary equation of photons has the stationary limit in the
region of drift velocities $(\vec{u}\vec{q}/\hbar \omega_q) < 1$,

\begin{equation}
\lim_{t\rightarrow \infty} N(\vec{q},t)=N(\vec{q})=
N(\vec{q},T_i) \left(1-\frac{\vec{u}\vec{q}}
{\hbar\omega_q}\right)^{-1}.
\end{equation}

As it follows from the equations (7), (11) and (12), for the drift
velocities $({\vec{u}\vec{q}}/{\hbar\omega_q})>1$ the distribution
function of photons grows exponentially with time. It corresponds 
to the generation of intrinsic photons by the increment of grow 
$\gamma_q$ and the amplification of photons introduced to the 
system from outside (stream of photons) by the coefficient of
amplification

\begin{equation}
\Gamma_q=\frac{\gamma_q}{c}=\frac{\beta}{c}\left(
\frac{\vec{u}\vec{q}}{\hbar\omega_q}-1\right)=
\frac{\beta}{c}\left[\frac{u}{c} \cos\alpha-1\right],
\end{equation}
here $\alpha=\widehat{\vec{u}\vec{q}}$ is the angle between drift 
velocity of the coupled by the mutual drag system 
``electron + photons'' (``the dressed electron'') or ``positron +
photons'' (``the dressed positron'') and the momentum of photon.

In the case of the propagation of strong electromagnetic wave 
and in the presence of external magnetic field, the current of 
electrons and positrons has the form:

\begin{equation}
\overrightarrow{j_{\pm}}=n e \overrightarrow{V^{\pm}}. 
\end{equation}
where $\overrightarrow{V^{\pm}}=\langle\overrightarrow{V^{\pm}}
(\varepsilon)\rangle$ is the average drift velocity of the carriers. 
Here

\begin{eqnarray}  
\overrightarrow{V^{\pm}}(\varepsilon) =\mp \frac{e~\Omega_{\pm}
(\varepsilon)}{m_c} \frac{\overrightarrow{E_{\mp}}\left(\omega_H^{\pm}/ 
\Omega_{\pm}(\varepsilon) \right) \vec{h}\vec{E}-\left(\omega_H^{\pm}/
\Omega_{\pm}(\varepsilon)\right)^2 \vec{h}(\vec{h}\vec{E})}
{\Omega_{\pm}^2-(\omega_H^{\pm})^2}, \\
\nonumber \vec{h}=\vec{H}/H, ~~ \Omega_{\pm}
\left(\langle \varepsilon \rangle \right)=\omega +i\nu^{\pm}
\left(\langle \varepsilon \rangle ,u\right), \\
\nonumber \nu_{ph}^{\pm}\left(\langle \varepsilon 
\rangle,u\right)=\nu_{ph}^{\pm}(T_i) \left(1- 
\frac{u\left( \langle \varepsilon \rangle \right)} 
{V^{\pm}\left(\langle \varepsilon \rangle \right)}\right). 
\end{eqnarray}

In the case of $\omega \rightarrow 0,$ and $\vec{E}
\Vert \vec{H}$ (or $H=0$) we have:

\begin{equation}
\overrightarrow{V}^{\pm} \left(\langle \varepsilon \rangle 
\right)=\frac{e \vec{E} \beta_c}{m_c \nu_{ph} \left(\langle 
\varepsilon \rangle,u \right) \beta_{ph,b}}=\frac{e\vec{E}}
{m \left(T_i,u \right)\beta_{ph,b}}.  
\end{equation}
Here $m=m_c \nu_{ph}\left(\langle \varepsilon \rangle,u\right) 
/ \beta_c$ is the mass of the coupled by the mutual drag system 
of carriers and photons and $\beta_{ph,b}=\beta_{ph}+\beta_b$.

Actually, the interacting system of heated electrons, positrons 
and photons in the stationary state $(\vec{u}\vec{q}/\hbar \omega_q) 
<1$ under external high electromagnetic, and the classically high 
or the quantizing magnetic fields as a result of their mutual drag 
has a cyclotron resonance with frequencies

\begin{equation}  
\omega_H^{\pm} \simeq \frac{e H}{m(T_i)c}=\frac{e H}
{\left(T_i/c^2 \right)c}.
\end{equation}

As it is seen from the equation (16), the resonance takes place with
frequencies of electromagnetic wave less than the collision 
frequencies of photons with carriers. The width of the resonance 
lines are defined by the expression 

\begin{equation}
\gamma=(3/2)\left[\left(\omega^2/ \beta_c\right)+\beta_{ph}+
\beta_b \right].
\end{equation}

In other words, as a result of the mutual drag, the electrons and
positrons become composite particles (the coupled system of 
``electron + photons'' and ``positron + photons'' i.e. so--called
``dressed'' by the photons ``quasi--electron'' or ``quasi--positron'' 
with an effective mass $m(T_i)$. In fact, we obtain the quasi--particle
with the electron's or positron's charge and the photon's mass 
 
\begin{eqnarray}  
m(T_i)=m(E,H)=T_i/c^2, ~~ \langle E \rangle=T_i=m(T_i)c^2.
\end{eqnarray}

Since $T_i$ and $T$ mean the average kinetic energy, then from 
the equations (13), (14) and (18), we obtain that the so--called 
``velocity of light'' in vacuum $c$ is the average velocity of 
photons in the equilibrium or stationary state at temperature $T$. 
The ``dressing'' of electrons and positrons in quantum 
electrodynamics was connected with the virtual absorption and 
emission of photons by electrons and positrons which occupies the 
given level of energy, i.e. with the finiteness of the lifetime 
of electrons and positrons or the natural width of the given 
energetic level. For the interacting system of electrons, holes 
and phonons in semiconductors, semimetals and gaseous plasmas, 
the analogous problem was solved in\cite{1}--\cite{5}.

As it follows from (7), (11), (12) and (13), under the external 
electric and magnetic fields and under stationary conditions, 
the relativistic factor enters the distribution function of 
photons in the first order $[1-(\vec{u}\vec{q}/\hbar \omega_q)]^{-1}$ 
instead of the form $[1-(v^2/c^2)]^{-1/2}$ in the relativistic 
electrodynamics. This is connected with the violation of time 
symmetry $(t \rightarrow -t)$ of equations in electrodynamics and, 
in common dynamics when there exists external fields. In this case 
the uniformity of the space and as a result, the law of the 
conservation of momentum is violated too. Since the external fields 
act continuously but not instantaneously we have a motion with
acceleration and the oscillatory regime is absent. The substitution
$t\rightarrow -t$ does not lead to the simple substitution
$\vec{v}\rightarrow -\vec{v}$, because the motions along and in the
opposite of the field direction are different and do not compensate 
each other.

The relativistic factor of the form $[1-(v^2/c^2)]^{-1}$ may appear 
only in the absence of the external field or in the weak external 
field in equilibrium or stationary conditions, by using the isotropic 
part of the distribution function of photons. Actually, by the 
separation of the stationary distribution function of photons (13) 
into the isotropic and anisotropic parts we have:

\begin{equation}  
N(q)=N_s(q)+N_{\alpha}(q) \\
\nonumber =N(q,T_i)\left(1-\frac{u^2}{c^2}
\cos^2\alpha\right)^{-1}+N(q,T_i)\frac{u}{c} 
\cos\alpha\left(1-\frac{u^2}{c^2}\cos^2\alpha\right)^{-1}.
\end{equation}

Since as a result of the mutual drag, the carriers and photons form 
the coupled system (complex) with a common drift velocity then under 
the conditions of strong (full) mutual drag $\alpha=0$ or $\pi$ and 
we obtain:

\begin{equation}
N_s(q)=N(q,T_i)\left[1-\frac{u^2}{c^2}\right]^{-1},
~~N_{\alpha}(q)=\left(\frac{u}{c} \cos\alpha \right) 
N(q,T_i)\left[1-\frac{u^2}{c^2}\right]^{-1}.  
\end{equation}

In the absence of external electric and magnetic fields, 
in the general case $u=u_0=const$ and we have:

\begin{equation}
N(\vec{q},\overrightarrow{u_0})=N(\vec{q},\vec{r}-
\overrightarrow{u_0}t,0)=\left\{ exp\left(\frac{
\hbar\omega_q^{\ast}}{T}\right) -1 \right\}^{-1},
\end{equation}
and $\hbar \omega_q^{\ast}=\hbar\omega_q-\overrightarrow{u_0}\vec{q}$. 
By the transition to the frame of reference drifting together with 
photons, we can obtain from (21) the equilibrium Planck's distribution 
function with temperature $T$. Let us discuss the main question now: 
may the presence of the relativistic factor of the first or the arbitrary
order lead to any singularities in physical phenomena or quantities? 
As it follows from the non-stationary solution of the distribution
function of photons (11) or (12), the Lorentz--Einstein theory
corresponds to uniform (equilibrium) case and must satisfy the 
stationary condition $v/c<1$! The case of $v=c$ is not included in 
their theory and therefore the conclusions of the Einstein's theory 
about being the rest mass of photons equal to zero and $c$ being the 
ultimate velocity of propagation of all types of interaction in nature 
do not have the real basis.

Actually, from general solution of the non--stationary kinetic equation
of the photons (11), by expanding the exponent into series near the point
$\vec{u}\vec{q}/\hbar \omega_q=1$ we have

\begin{eqnarray}
N(\vec{q},t)=\left\{ N_0(\vec{q},T)+\frac{\beta N(\vec{q},T_i)}
{\gamma_q}\right\} \left\{1+\gamma_q t+\frac{1}{2}(\gamma_q t)^2
+...\right\}- \frac{\beta N(\vec{q},T_i)}{\gamma_q} \\
\nonumber =\left\{N_0(\vec{q},T)(1+\gamma_q t)+N(\vec{q},T_i)
\beta t\right\} +\frac{1}{2}\left\{N_0(\vec{q},T)
(\gamma_q t)^2+\beta\gamma_q N(\vec{q},T_i) t^2 \right\}.
\end{eqnarray}
At the point $\gamma_q=0$ we have

\begin{equation}  
N(\vec{q},0)=\lim_{\gamma_q\rightarrow
0}N(q,t)=N_0(\vec{q},T)+N(q,T_i)\beta t.
\end{equation}

As it follows from the expressions (24) and (25), at the point
$\vec{q}\vec{u}/ \hbar \omega_q=1$, i.e. at the point $u=c$ the
distribution function of photons is non-stationary and grows
linearly with time and the singularity is canceled!

Since the Einstein's theory is a stationary one, it is not applicable 
to the non--stationary conditions, namely to the region of drift
velocities $v\geq c$ or $u\geq c$. For the drift velocities 
$v\geq c$ or $u\geq c$ the theory must be non--stationary. The Einstein's
theory is a one mode theory and that is why it must be obtained from the
many particle (or many mode) theory by the limiting transition to the one
mode case. The effect of coupling of charged carriers and photons as a
result of their mutual drag is a common phenomenon. It is a reaction of
the system on action of external fields for the conservation of the
stationary state of system (the analog of ``self--conservation'' in
biology). Thus we have found the ``dressing'' mechanism of charge 
carriers used earlier in quantum electrodynamics.

\section{Conclusions}

As a result of analyses of the phenomena of coupled system of charge 
carriers and photons, by using the equations (7),(10)-(13) and (20) 
we reach the following conclusions:

{\bf 1.} As it follows from the non--stationary distribution function 
of photons the Lorentz-Einstein theory corresponds uniform space 
(equilibrium) case and it must satisfy the stationary condition 
$v<c$! The case of $v=c$ is not included in their theory. At the 
point $u=c$ (i.e. $\gamma_q=0$), the distribution function of 
photons (23) is non--stationary, does not have any singularity and 
grows linearly with time.

For this reason, the conclusions of the theory of relativity that 
the rest mass of photons is zero and $c$ is the ultimate velocity 
for the propagation of all types of interactions in nature do not 
have the real basis.

{\bf 2.} The Einstein's theory was a one mode theory and that is why must
be derived from the many body (mode) theory by limiting transition to the
one mode case at $v<c$. For the $v>c$ or ($u>c$) cases the theory must 
be non--stationary. As we say the mutual drag leads to the formation of
``quasi--particles'', the electron or positron ``dressed'' by photons. 
The average energy in stationary state $u<c$ for the one mode case is:

\begin{eqnarray}  
\langle \varepsilon \rangle=\langle \hbar \omega \rangle 
\langle N_i(q,T_i) \rangle \approx \frac{T_i \langle N(\omega,T_i) 
\rangle}{1-u^2/c^2} \\
\nonumber =\frac{T \langle N(\omega,T_i)\rangle}{1-u^2/c^2}
\frac{T_i}{T}=\frac{M_0 c^2}{1-u^2/c^2}
\left(\frac{T_i}T \right)^2.
\end{eqnarray}
The mass of the heated photons for one mode is:

\begin{equation}
M_i=\frac{M_0}{1-u^2/c^2}\left(\frac{T_i}T\right)^2.
\end{equation}
The mass of one heated photon is:

\begin{equation}
m=\frac{M_i}{\langle N (\omega,T_i) \rangle}=m_0 
\frac{T_i}T\left(1-\frac{u^2}{c^2}\right)^{-1}.
\end{equation}
In the absence of heating, the average energy is:

\begin{eqnarray}
\langle \varepsilon \rangle =Mc^2=\langle N_0(\omega,T) \rangle 
\frac {T}{1-u^2/c^2}=M_0c^2\frac {1}{1-u^2/c^2} \\
\nonumber =m_0c^2\frac{\langle N_0(\omega,T)\rangle}{1-u^2/c^2}, ~~
m=\frac{m_0}{1-u^2/c^2},
\end{eqnarray}  
where $m_0=M_0/ \langle N_0(\omega,T)\rangle \simeq T/c^2$ is 
the rest mass of photon, i.e. the mass of photon in the frame 
of reference which is drifted with photons at temperature $T$, 
and $\langle N \rangle$ is the concentration of photons for 
one mode.

{\bf 3.} As it follows from (26) under high electric and magnetic fields 
for drift velocities $u<c$, the energy (or the mass) of photons for 
one mode grows as a result of the mutual drag, as well as the heating 
of the carriers and photons.

{\bf 4.} As it follows from (7)-(13) under the external electric and 
magnetic fields, the relativistic factor enters the expressions of 
the distribution function and other physical quantities in the 
first order as $\left(1-u^2/c^2\right)^{-1}$, instead of 
$\left(1-v^2/c^2\right)^{-1/2}$ in Einstein's theory. This together
with the conclusions 1 and 2 solves the main problem of super--luminal
particles named--tachyons, because in our theory the imaginaricy of
the mass of tachyons is liquidated.

{\bf 5.} There is an opinion that the original conception about tachyons, 
as individual particles such as electrons, protons and etc. is not
correct and the tachyons in such understanding is absent\cite{11}. 
{\bf Our investigations show that the ordinary particles such as 
electrons, positrons and also like photons may stand as 
super--luminal under high external fields in the condition of 
mutual drag}. Also there is an opinion that the tachyons as an 
elementary excitations (quasi--particles) there exists widely in 
complex systems which loss its stability and made an phase 
transition to a stable state\cite{11}. In origin the tachyons, in 
general, were considered only in amplifying mediums\cite{12}--\cite{15}. 
{\bf As it is seen from our investigations actually for drift
velocities greater than the speed of light, the super--luminal 
particles generate or amplify the electromagnetic waves (photons) 
and the super--luminal particles are placed in the regime of
generation or amplification independently of the type of medium 
(see also\cite{16})}. As it will be shown in a special report,
in general all elementary excitations including so--called 
``elementary particles'' are quasi-particles.

{\bf 6.} At the point $u=c$ the angle $\alpha$ between $\vec{u}$ and 
$\vec{q}$ is equal to $\pi/2$ and we have the condition that the
electromagnetic wave becomes free and is emitted. Thus the point 
$u=c$ is the point of transition of the system from the absorption 
regime to the regime of emission of electromagnetic waves (photons).

{\bf 7.} It is shown that the relativistic expression for the deceleration 
of time really takes place for the relaxation time of carriers and
photons, for the life--time of carriers and also for the period of
electromagnetic oscillations. Actually, in dynamics and electrodynamics
the time enters the relations as a parameter, but not as a free 
coordinate and for this reason the Einstein's relation is impossible 
to apply to the time. Since $\tau^{-1} \sim N_i(q,T_i)$ we have 

\begin{equation}
\tau_i \approx \tau_0(1-u^2/c^2)(T/T_i),~~ l_i=u\tau_i=
l_0(1-u^2/c^2)(T/T_i).
\end{equation}
If $T_i=T$ we have: 

\begin{equation}
\tau_i\approx \tau_0(1-u^2/c^2),~~ l=l_0(1-u^2/c^2).
\end{equation}

{\bf 8.} It is shown that the so--called ``speed of light'' in vacuum 
$c$, like the velocity of sound for phonons, is an average velocity 
of photons in the ground state.

{\bf 9.} As it follows from (7), (10), (11) and (13) at the point 
$u=c$ the distribution function of photons is an isotropic one and 
the anisotropy part of the photon distribution function at this 
point is zero. It means that the ground and the stationary states 
of electromagnetic field is spherically symmetric (this question 
will be discussed in a special report). At the point $u=c$ the 
rate of the stimulated absorption and emission are equal to each 
other and there is only spontaneous emission of photons and 
dissipation is absent.

{\bf 10.} It is shown that the demand for the invariance of the 
Maxwell equations or the laws of Physics in all inertial frames of
reference is not correct. It is equivalent to the demand of 
invariance of laws of Physics in the uniform and non--uniform 
spaces or to the demand of equivalency of physical laws in cases 
of the presence and absence of external field (force).

{\bf 11.} It is shown that the presence of the second inertial 
frame of reference moving with a constant drift velocity relative 
to the first one may be a result of the presence of space 
non--uniformity or the uniform external field.

Actually in the uniform space because of the equivalency of all 
the points of space, it is impossible to have simultaneously two 
or more inertial frames of reference with different constant drift
velocities with respect to each other. Because it will lead to the
violation of the space uniformity, i.e. the change of the distance
$r_{12}$ between the initial points of that frames of reference
($r_{12}\neq const.$). In the uniform space, for the conservation 
of the space uniformity during the motion it is necessary that 
the motion of all points of the space must be the same constant 
velocity (in the absence of the external field), or with the constant
acceleration (in the presence of the uniform force). For both cases 
the frames of reference, connected with the different points of the 
space, do not have the motion relative to each other without the 
violation of the space uniformity or the uniformity of the external
force. Since all points of the space in both cases are placed on 
the same conditions and are equivalent. To choose two frames of 
reference moving with constant drift velocity with respect to each 
other, it is necessary to have the space, which consists of two or 
more sub--spaces. All points of the first sub--space are at rest or 
have the motion with the constant velocity $v$ and all points of the 
second sub--space move with a constant acceleration. The first of 
them is inertial, but the second accelerates and ,for this reason, 
the demand of equivalency of laws of Physics in that two frames of
reference are equivalent to the demand of the equivalency of the 
first and the second Newton's laws. This demand is absurd, of course!

{\bf 12.} As it follows from (7)--(12) the demand of the equivalency 
of the laws of Physics in all inertial frames of reference for the 
photons is equivalent to the demand of the equivalency of the Planck's
equilibrium distribution function to the drifted Planck's distribution
function.

{\bf 13.} It is shown that at the external electric fields $E<H$, the
drift velocity of carriers $u<c$ and the energy received from external
field is accumulated by the coupling of carriers with photons, as a 
result of the mutual drag (as a result of the construction of structure 
by the mechanism of ``dressing of carriers with photons'') and stationary
state is conserved. In this region of drift velocities the absorption
is greater than the emission. Under the condition $E>H$ i.e. at the 
drift velocities $u>c$, the generation and amplification dominates and
the number of photons grows exponentially with the time. {\bf The
violation of the stationary state begins from the point $u=c$ and from
this point it begins that the transition of the carriers ``dressed by
photons'' to the following stationary state by the reactive emission of 
photons back}. In the region of drift velocities $u>c$, the 
anisotropic part of the photon distribution function is much more 
than the isotropic one. {\bf Thus in our investigation the mechanism 
of the transition of particles to the following stationary state is
obtained.}

{\bf 14.} It is shown that for the drift velocities $u>c$ so--called
energy (or mass) for one mode is:

\begin{eqnarray}
\langle \varepsilon \rangle =\left(\frac{T_i}{T} \right)^2
\frac{T \langle N(q,T) \rangle}{(u/c)-1}
\left\{\left[\exp (\gamma_q t) -1 \right]+ \frac{T}{T_i}
\exp(\gamma_q t) \right\} \\ 
\nonumber  =\frac{M_0c^2}{(u/c)-1}\left(\frac{T_i}{T} \right)^2
\left\{\left [\exp(\gamma_q t)-1 \right]+\frac{T}{T_i}
\exp(\gamma_q t)\right\},
\end{eqnarray}
or the mass for one mode is:

\begin{equation}
M=\frac{M_0}{(u/c)-1} \left\{\left[\exp(\gamma_q t)-1 \right] 
+ \frac{m_0}{m(T_i)} exp(\gamma_q t) \right\}.
\end{equation}  

In the absence of heating

\begin{equation}
M=\frac{M_0}{(u/c)-1} \left[ 2 \exp(\gamma_q t)-1 \right] 
\approx \frac{2M_0}{(u/c)-1} \exp(\gamma_q t).  
\end{equation}
As it follows from these equations the mass of photons for one mode 
in the region of drift velocities $u>c$ grows exponentially in time.

The case considered by Einstein may correspond to the case when the 
first of the chosen frame of reference is connected with the photons 
and drifted together with them and the second is connected with the 
charged carriers (electrons or positrons) drifted with constant 
velocity relative to first one. Here the photons (electromagnetic
field) play the role of media and the charged carriers drifted relative 
to that media. Thus the system must consists of three subsystems. 
Actually, in electrodynamics the space (or the system) is consists of
three sub--space (or subsystems): the sub--space of negatively charged
carriers, sub--space of positively charged carriers and the other
sub--space (or space) of photons. For this reason, the electrodynamic
space is non--uniform initially, because the point of space where
negatively charge carrier is placed is not equivalent to the point
of space where the positively charged carrier is placed and both are
not equivalent to the point where the photon is placed. The presence 
of the two type of charge leads to the presence of so--called Lorentz
force. The condition of stationarity of the ground state is the equality
of this force to zero $F=0$! This condition corresponds to the
annihilation of charge carriers with production of photons, i.e. 
production of free electromagnetic field without charges. It means that
the space of photons (i.e. free electromagnetic field) can decay to the 
two sub--spaces: the spaces of the negative and positive charges and also 
two sub--spaces of negative and positive charges can produce the space 
of photons (the electromagnetic space or media).

As it follows from our investigations, the case studied by Lorentz 
and Einstein corresponds to the case of the presence of weak external
field when the heating of carriers and photons is absent and there is
only their mutual drag. Also they considered the case of one type of
charge carriers. In the presence of mutual drag of the electrons and
photons, the distribution function of photons $N(q)$ has the form of 
Planck's function displaced with a constant drift velocity and as a 
result with the renormalized frequency of emission $\omega_{em}=
\omega_q^{\ast}=\omega_q-\vec{u}\vec{q}/ \hbar$. For the drift
velocities $u<c$, $\omega_{em}$ decreases with the increasing drift
velocity $u$ because the stationary distribution function of photons 
has the form (23) or (13).

Thus in the second frame of reference, charge carrier emits or 
absorbs photon with a frequency $\omega_{em}=\omega_{obs.} 
\left[1- \vec{u}\vec{q}/ \hbar \omega_{obs.}\right]$. In the one 
mode case for the observed frequency we have $\omega_{obs.}= 
\overline{\omega}=T_i/ \hbar=cq/ \hbar$. In other words

\begin{equation}
\omega_{obs.}=\omega_{em}\left(1-\frac{u}{c}\cos \alpha\right)^{-1},
~~\lambda_{obs.}=\lambda_{em}\left(1-\frac{u}{c}\cos \alpha \right).
\end{equation}

For the case when the both frames of reference are assumed to be 
inertial one, i.e. they move along one line (along the $x$--axis) 
$cos\alpha =1$ and we have

\begin{equation}
\omega_{obs.}=\omega_{em}\left(1-\frac{u}{c}\right)^{-1},
~~\lambda_{obs.}=\lambda_{em}\left(1-\frac{u}{c}\right). 
\end{equation}

As it follows from these equations, the Doppler effect is also a 
result of the mutual drag of carriers and photons. Actually the source 
of the emission (charge carrier) drifts relative to frame of reference
connected with the photon (observer) with the drift velocity $\vec{u}$. 
By increasing of drift velocity $\vec{u}$ the distance between the 
source and the detector is increased too and ,as a result, the observed
frequency of photons is increased or the wavelength is decreased.
In the opposite case, the wavelength is increased. In the region 
of $u<c$ the source (the charge carrier) moves slower than the detector
(photon) and as a result the distance between them is decreased and the
wavelength of observed photons (light) is increased. Thus the mutual 
drag of electrons and photons in the region $u<c$ leads to decreasing 
the frequency of emission or absorption (see ``Low frequency cyclotron
resonance for the phonons'' \cite{2,5,8}). It means that the frequency 
of the observed light (photons) is increased.

{\bf 15.} As it follows from the present consideration the so--called
relativistic phenomena and the Doppler effect are the same ones and 
are a result of the mutual drag of interacting system of carriers and
photons under external field. The case was considered by Lorentz and
Einstein corresponds to the case, when the charge carriers are drifted 
under electromagnetic field at the conditions of the mutual drag of
carriers and photons in the absence of their heating by the field.

The distinction between the results of the theory of relativity and 
the Doppler effect is related with the fact that the Doppler effect 
deals with the total stationary distribution function whereas the 
theory of relativity deals only with its isotropic part. In other 
words, the Doppler effect is obtained as a result of taking into 
account the violation of the time--symmetry at the external field.

\section{The Superluminal Motion}

In 60's and at the beginning of 70's years the problem of possibility 
of superluminal motion was a subject of wide spread discussions\cite{17}. 
The hypothesis on the existence of superluminal particles--tachyons--was
introduced in \cite{15}-\cite{20}. Although, the  special theory of 
relativity does not strictly outlaw tachyons. in order to remain within
the demands of the special theory of relativity, it was necessary to
ascribe the imaginary mass $(im_o)$ for tachyons. But, the imaginary 
mass of tachyons does not call a great anxiety. More complexities were
connected with the principle of causality. There was witty but
unsuccessful attempts to reconcile the existence of tachyons with the
causality principle\cite{15,19}. At that time there was no experimental
evidence for tachyons. Therefore, the interest in tachyons disappeared 
slowly.
 
In 1965, a physical object moving with superluminal velocity was
realized. It was a pulse of light with stationary semblance moving
through the amplifiered laser with a velocity greater than the speed 
of light in vacuum\cite{20}-\cite{22}. The propagation velocity of 
the light pulse through the amplifier exceeded the vacuum speed of   
light by 6--9 times.

In 1970, Garrett and McCamber\cite{23} asserted that the concept of group 
velocity is meaningful only in an absorptive medium. The conclusion was 
based on the simulations  of Gaussian pulses propagating in a medium 
which has resonance centered on the narrow spectrum of the pulse. They 
found that the peak of pulse can travel the medium with apparent
superluminal velocities and even leaves the medium before the original
peak enters the medium, i.e. negative velocity. This took place even the
pulse maintained a semblance of its original temporal profile, albeit
attenuated, allowing for an unambiguous assessment of travel time 
of pulse. They explained this interesting event which seemed to be 
in contradiction with the theory of special relativity and the 
principle of causality, in terms of pulse reshaping where the 
latter portion of the pulse is preferentially attenuated so that 
the pulse appeared to move forward with a excessive speed, which 
agreed with the group velocity.
 
In 1974, a paper was published\cite{24} in which the superluminal 
motions were connected with the motion of excitations in unstable 
media which loses its stability and makes phase transition into 
more stable state. As an example of the unstable media mentioned, 
the authors discussed in a paper\cite{25} too. The idea about 
superluminal motions in unstable media were developed in \cite{25}. 
As it follows from our theory, electrons and positrons generate 
or amplify the photons with superluminal velocities in 
nonstationary or unstable regime.

In 1982, S. Chu and S. Wong\cite{26} confirmed the predictions of 
Garret and Mc Camber by measuring the transmission time of laser 
pulse turned to a resonance in GaP:N. In particular, the physical
relevance of group velocity was affirmed in the superluminal and 
negative cases.

At Berkeley, Steinberg, Kwiat and Chiao\cite{27,28} have experimentally
showed that individual photons penetrate into optical tunnel barrier 
with a group velocity considerably greater than the vacuum speed of 
light. In the papers\cite{29,30}, R. Chiao et al. reported that in
simulations of pulse propagation  in amplifier media, near gain 
resonances similar exotic behavior is expected. Several experiments on
photonic tunneling observed in microwave and optical regions showed 
the superluminal signal and energy velocities violating Einstein
causality. Superluminal group velocities have been experimentally 
observed for propagation through an absorbing medium\cite{26}, for
microwave pulses\cite{31,33} and for light transmitted through a
dielectric medium\cite{27,28,29,34}. These experiments demonstrates 
the superluminal nature of the tunneling process. Tunelling is one of 
the most mysterious phenomena from quantum mechanical point of view. 
Yet it is one of the most basic and important process in nature. It may
seem that all that we want to know about the tunneling is now well
understood as given in the present mature of quantum theory. However,
there remains an open problem concerning the mechanism and duration of 
the tunneling process, i.e. the question: ``How long does it take for the
particle to tunnel across the barrier?". This question first addressed 
in 1930's\cite{35} is still the subject of much controversy, since the
numerous theories contradict each other in their predictions for the
tunneling time. Actually some of them as most notable Wigner's theory
predict that this time should be superluminal\cite{36}, but others 
predict that it should be subluminal\cite{37,38}. Thus, the first answer
that the group delay (also known as the ``phase time") can  in certain
limits be paradoxical small, implying barrier transversal at a speed
greater than that of light in vacuum\cite{35,36}.

Apart from its fundamental interest, a correct theoretical and 
experimental solution of this problem is important from general point 
of view and for the determining the speed of devices which are based 
on tunneling. Prof. R. Chiao\cite{39} states that ``... The many
manifestations of tunneling and the many applications to devices 
strongly motivated us to examine the unsolved tunneling time problem
experimentally." It is very important to attack to state precisely at 
the outset the operational definition of the quantity being measured. 
For their experiments, the ``tunneling time" was that type of quantity.

The photonic tunneling experiments in microwave and optical regions 
have revealed superluminal signal and energy velocities\cite{27,28,31}. 
There is a opinion that the photonic tunneling violates the Einstein
causality\cite{40}. The  measured velocity tunneled signal was $4.7c$ 
for the microwave\cite{28} and $1.7c$ for single photon\cite{27,28} 
experiments.

Recent experiments were carried out by using three examples of photonic
barriers: an undersized wave guide between two normal guides, a periodic
dielectric heterostructure (often called a one dimensional photonic
lattice) and double prism with a gap of rarer refractive index acting as 
a photonic barrier. Latter set up described as frustrated total internal
reflection.

Let us now present the results of experiments carried out in 1965 by 
Bassov et al. which are very important in theoretical point of view. 
At the beginning of 60's arised the problem of receiving the high 
energy pulses in time interval about 1 nanosecond. For this purpose, 
a short pulse of light is obtained with the help of so--called
giving generator and amplified by an amplifier laser\cite{12,21,22}. 
The schematic description of the experimental set up is as follows: 
The pulse of light obtained from the giving generator splits into two
parts. First, more powerful part propagated through the amplifier and 
the second part propagated through the air and played the role of the
registration target. Both of the pulses were received by the emission 
receivers. The signals of the receivers were sent to the oscilloscope 
for visual observation. There was an expectation that the pulse 
propagated through the air must have a speed greater than that of the 
pulse propagated through the amplifier. Moreover, there was a prior
expectation that together with the increasing of the amplitude of the
pulse propagated through the amplifier, its form must also change. The
real results of the experiment were surprising and leaded to some 
confusions. The pulse propagated through the amplifier did not change 
its form and more paradoxical propagated with a velocity 6--9 times
greater than that of the light in vacuum $c$. 

Let us now try to understand what was happened in the photonic experiments
carried out by the propagation of light pulses through the amplifier and 
through the photonic barriers for determination of tunneling time and to
answer the question: ``Does the motion of photons across the amplifier or
across the photonic barriers subluminal or superluminal?". Our theory
answers this question: This motion must be a superluminal and it does not
affront to the theory of special relativity and causality. Because, as it
follows from the theory considered in the previous pages of the present
paper, the theory of special relativity is applicable for the stationary
(equilibrium) state in the region of drift velocities $u<c$. In the 
region of drift velocities $u>c$, the distribution function of photons is
non--stationary and grows exponentially in time. At the point $u=c$ (at
light instability threshold--LIT) as it follows from (20), the 
distribution function of photons is spherically symmetric one and it does
not consist of any singularity and grows linearly in time. At this point
the stimulated emission and absorption rates are the same and there is
only spontaneous emission of photons at high external field. Moreover, 
the dissipation is absent and it is the point of stationary state. At the
condition $u=c$, coupled system of carriers and photons emits all the
power gained from the external field as photons (or electromagnetic
waves) and as a result conserves the its stationary state. At this point
the electrical current is constant and so the state is dynamically
stationary one. Only from this point, the emission of electromagnetic
waves by charge carriers begins under external electric field. As it
follows from equations (7)--(9) only the part of photons with drift
velocities $u<c$ are damped, the other part of photons with drift
velocities $u>c$ are generated or amplified at external fields so they 
are undamped in the absence of external fields. As it was shown 
in our theory the so--called the speed of light in vacuum $c$ is the
average (by the frequencies) velocity of the pulse, or the velocity of
photons with a frequency $\omega=\omega_{peak}=\overline{\omega}$, in 
the ground state. For other stationary states $c$ is the average 
thermal velocity $\overline{v}=\sqrt{\overline{\varepsilon}/m} 
\approx \sqrt{T_i/m}$ of pulse Thus, only the subluminal (or subthermal) 
photons, $u < c$, at Planck's distribution function are interacting 
with the medium, but the superluminal photons, $u \gg c$, transmits
into the medium freely without loss and the form of the pulse is 
conserved. This phenomena is independent of the type and the width of
photonic barrier. This result  represents a fundamental behavior of all
the wave packets or light pulses in nature. Moreover, it shows that the
tunneling processes are superluminal initially. There is no mysterious 
in the tunneling; the superthermal particles with energies (or 
frequencies) greater than the height of the barrier transmits the 
barrier region freely, because there is no barrier for them. Only the
subthermal photons, frequencies with $0 < \omega < \overline{\omega}$,
are reflected or damped through the process of interaction with
barrier. It is well known there is no absolutely opaque barrier for all
frequency region. The mediums which is opaque for the visible light is
transparent for the region of Roentgen and Gamma rays. In other words, 
the medium which is opaque for low frequencies becames transparent for 
the high frequencies of photons. We found that the critical frequency is
the average frequency of pulse or wave packet and obtained from the
average energy or temperature of pulse. In the case when the light pulse
propagates in vacuum or through in air, then the pulse propagates as a
unit system, as a particle with energy (or frequency) 
$\hbar \overline{\omega}$ (or $\overline{\omega}$) and it does not 
change its form. All points of pulse propagates with the same velocity 
of the peak of pulse. In this case, the pulse propagates without
interaction and for that reason conserves its form during the propagation.

In the case of the propagation of light pulse through the barrier, the
pulse interact with the barrier and splits into two pulse. The pulse 
of subluminal (or subthermal, damped), photons with drift velocities 
$u < \overline{v} \equiv c$ and superluminal (or superthermal, undamped) 
pulse of photons. While the subluminal pulse is reflected from the
photonic barrier or damped, the superluminal pulse is propagates 
through the barrier without interaction and conserves its form. This
theoretical consideration is confirmed by the photonic
experiments\cite{12,21,22,28,31,34}. 

In the case of the propagation of light pulse in external electric and
magnetic (or electromagnetic) fields, for example, through an amplified
laser (see [12]), the pulse splits into two pulse, superluminal and
subluminal. This can be seen visually. As these pulses propagates through 
the amplifier, the subluminal one is damped and the superluminal one is
amplified. If the time interval $\Delta t$ between the two acts of
photons gaining energy from the external field is greater than the
relaxation time of photons $\tau_{ph,ph}$ in the photon--photon 
scattering event, the pulse propagates without changing its form, but 
its amplitude must be increased. On the other hand, if 
$\Delta t < \tau_{ph,ph}$ then photons does not enough time to establish
the equilibrium distribution of pulse with effective temperature, $T_p$. 
As a result, the form of the pulse should be changed and we have the case 
while the amplitude of the lower frequency part of the pulse will be
decreased in time, the amplitude of the higher frequency (superluminal) 
part of the pulse will be amplified and finally we must have the high
frequency half part of the initial pulse with amplitude growing in 
time\cite{12}.

Experiments with propagation of light pulse through the amplifying system
confirms the theoretical picture stated above\cite{12,21,22}.
\\
\\
\\
{\bf Acknowledgments}
\\
This work is partly supported by NATO--TUBITAK.

{\bf PS:} The original version of this paper was reported in $2^{nd}$
Cracow--Clausthal Workshop  ``Extensions of Quantum Physics''
Cracow, 12--14 October 2000 as an invited report.

\end{document}